\documentclass[aps,prb,twocolumn,10pt,superscriptaddress,reprint,floatfix,nofootinbib]{revtex4-2}
\usepackage{indentfirst}
\usepackage[utf8]{inputenc}
\usepackage{booktabs}
\usepackage{amsmath}
\usepackage{multirow}
\usepackage{geometry}
\usepackage{xcolor}
\newgeometry{vmargin={13mm}, hmargin={12mm,16mm}}   %% set the margins
\usepackage{amsmath,amssymb,stmaryrd}   
\usepackage{xcolor}
\usepackage{upgreek}
\usepackage{parskip}
\usepackage{xfrac} 
\usepackage{graphicx}
\usepackage{lipsum,babel}
\usepackage{dcolumn}
\usepackage{bm}
\usepackage{siunitx}
\usepackage{float}
\usepackage{esvect}
\usepackage{gensymb}
\usepackage{commath}
\usepackage{natbib}
\usepackage{mathtools}  
\usepackage{diffcoeff}
\usepackage[version=4]{mhchem}
\usepackage{chemformula}
\usepackage[colorlinks=true,linkcolor=blue]{hyperref}%
\begin{document}

\preprint{APS/123-QED}

\title{Non-relativistic linear Edelstein effect in helical \ch{EuIn2As2}}

\author{Nayra A. \'Alvarez Pari}
\affiliation{Institut für Physik, Johannes Gutenberg Universität, D-55099 Mainz, Germany}
\author{Rodrigo Jaeschke-Ubiergo}
\affiliation{Institut für Physik, Johannes Gutenberg Universität, D-55099 Mainz, Germany}
\author{Atasi Chakraborty}
\affiliation{Institut für Physik, Johannes Gutenberg Universität, D-55099 Mainz, Germany}
\author{Libor  \v{S}mejkal}
\affiliation{Max Plank Institute for the Physics of Complex Systems, N\"{o}thnitzer Str. 38, 01187 Dresden, Germany}
\affiliation{Max Planck Institute for Chemical Physics of Solids, N\"othnitzer Str. 40, 01187 Dresden, Germany}
\affiliation{Institut für Physik, Johannes Gutenberg Universität, D-55099 Mainz, Germany}
\affiliation{Institute of Physics, Academy of Sciences of the Czech Republic, Cukrovarnick\'{a} 10, 162 00 Praha 6, Czech Republic}
\author{Jairo Sinova}
\affiliation{Institut für Physik, Johannes Gutenberg Universität, D-55099 Mainz, Germany}
\affiliation {Department of Physics, Texas A \& M University, College Station, Texas 77843-4242, USA}

\date{\today}

\begin{abstract}

Motivated by the ongoing interest in understanding the actual magnetic ground state of the promising axion insulator candidate \ch{EuIn2As2}, we present here a spin symmetry analysis and \textit{ab initio} calculations, aiming to identify specific exchange-dominated physics that could offer insights into the current debate. We investigate two non-collinear coplanar magnetic orders reported in this compound: the helical and broken-helical phases. Our symmetry analysis shows that magnetic-exchange alone results in the formation of an out-of-plane odd-parity wave order in momentum space with a single un-polarized nodal plane in both phases. Additionally, we identify an in-plane g-wave order that emerges exclusively in the broken-helical phase, providing a distinguishing feature for this phase. Furthermore, we report a non-relativistic Edelstein effect with a distinct out-of-plane polarized spin density that dominates over spin-orbit coupling effects. Our \textit{ab initio} calculations reveal a significant contrast in the magnitude of this effect between both phases, which could serve as a mean to identify the magnetic transition and distinguish them from other magnetic ground states proposed for this compound.

\end{abstract}
\maketitle

\section{Introduction}
Understanding the intricate interplay between magnetic and topological properties of quantum materials is a fundamental challenge in condensed matter physics. The mutual influence is particularly evident in materials where electronic and magnetic degrees of freedom are coupled, as magnetic order can alter the energy dispersion and lead to unique topological features. In this context, the \ch{Eu}-based 122 compounds have attracted considerable interest due to their intrinsic
spin-orbit coupling (SOC) arising from their strongly localized $f$-orbitals, which can lead to distinctive electronic, magnetic, and topological properties in the band structures. The realization of various quantum phases within this family, including exotic topological surface states \cite{li2019dirac,sato2020signature,pierantozzi2022evidence,pari2024strain,zhao2024hybrid}, colossal magnetoresistance \cite{wang2021colossal,krebber2023colossal,luo2023colossal}, the axion electrodynamics \cite{xu2019higher, ma2020emergence}, and the recently altermagnetic phase \cite{cuono2023ab} make these compounds a promising magnetic topological platform for future spintronic applications.

\begin{figure}[h]
	\includegraphics[width=0.48\textwidth]{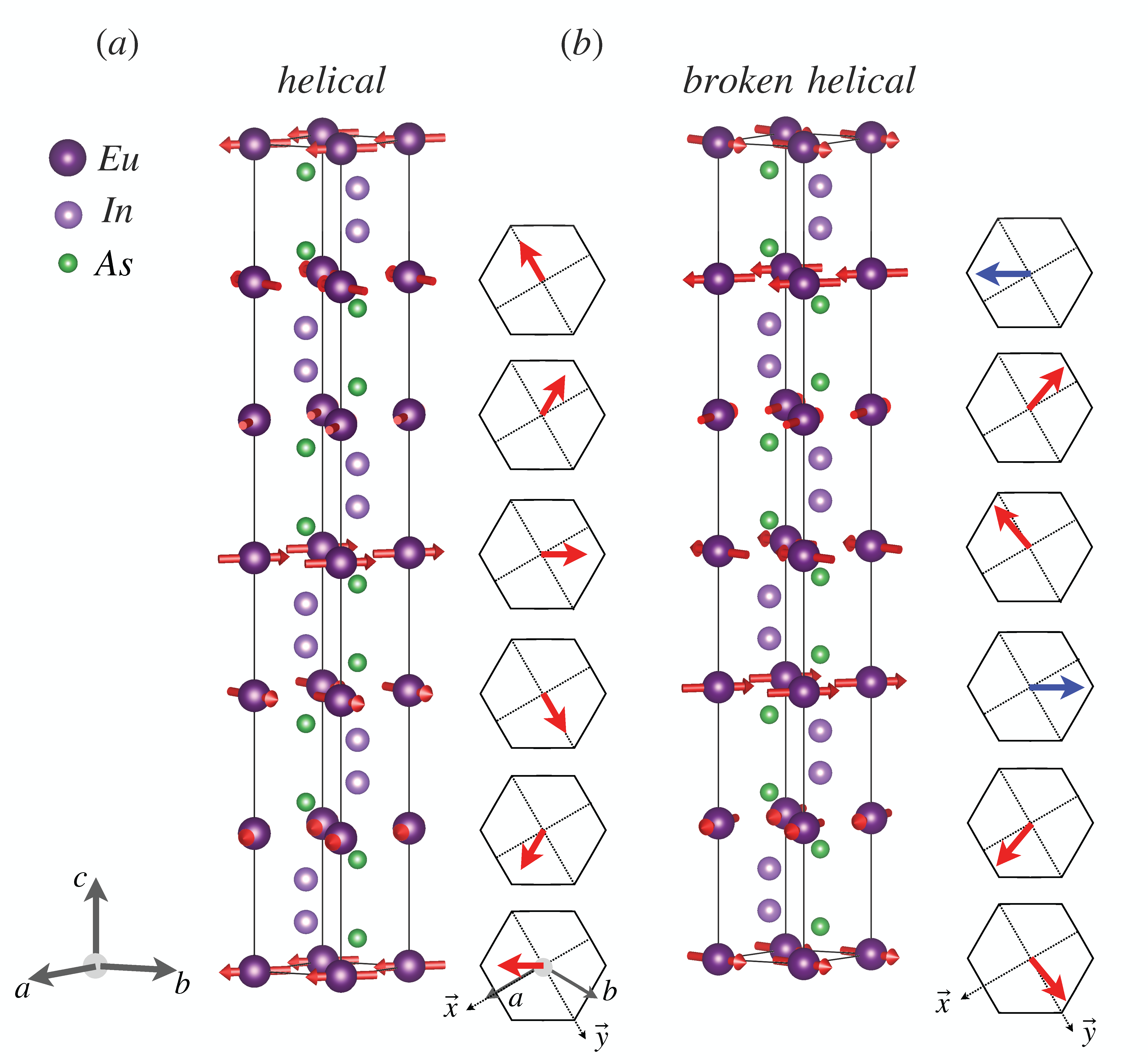}
	\caption{Non-collinear coplanar magnetic textures of \ch{EuIn2As2}. (a) The helical phase, with neighboring magnetic layers connected by $60\degree$ rotation along $\hat{z}$. (b) The broken-helical phase, with two symmetry-independent \ch{Eu} magnetic sub-lattices. The red-red spin sublattices are connected by $80\degree$, whereas the red-blue by $130\degree$. For both phases, a tripling of the unit cell along $\mathbf{c}=c~\mathbf{\hat{z}}$ is assumed \cite{riberolles2021magnetic,soh2023understanding}. }
 \label{struct}
\end{figure}

The rare-earth compound \ch{EuIn2As2} has particularly caught the spotlight within the 122 family after first-principles studies have proposed it as a candidate axion insulator, assuming a simple antiferromagnetic collinear order \cite{xu2019higher, zhang2020plane}. However, this initial theoretical prediction was quickly challenged by experimental studies proposing more complex magnetic structures, from commensurate \cite{riberolles2021magnetic, soh2023understanding,donoway2024multimodal} to incommensurate \cite{takeda2024incommensurate,gen2024incommensurate} magnetic phases. In this study, we focus on two commesurate coplanar non-collinear magnetic configurations %The magnetic structures 
% suggest coplanar non-collinear magnetic structures
-- namely, the helical and broken-helical phases -- reported by Riberolles et al. and Soh et al. ~\cite{riberolles2021magnetic,soh2023understanding}. The helical phase emerges at  $T_{\mathrm{N_1}}=17.6(2) \ \mathrm{K}$, with adjacent magnetic layers connected by $60\degree$ rotation along $\hat{z}$, as illustrated in Fig.\ref{struct}(a). At lower temperatures $T_{\mathrm{N_2}}=16.2(1) \ \mathrm{K}$, the broken-helical phase emerges, with a magnetic configuration involving two symmetry-independent Eu magnetic sub-lattices, highlighted in red and blue in Fig.\ref{struct}(b).  While previous studies have primarily focused on examining the magnetic space group symmetries of these phases -- essentially due to the compound's strong spin-orbit coupling -- we adopt a non-relativistic approach by analyzing the spin symmetries \cite{litvin1974spin, litvin1977spin}, aiming to properly describe the effects of magnetic exchange on the electronic band structure. 
%Additionally, recent first-principle studies have revealed non-relativistic spin splittings in the collinear magnetic configuration \cite{cuono2023ab}. Therefore, studying this compound employing spin symmetries \cite{litvin1974spin, litvin1977spin} could offer valuable information on the effects of magnetic exchange on the electronic band structure.

\begin{figure*}
	\includegraphics[width=.95\textwidth]{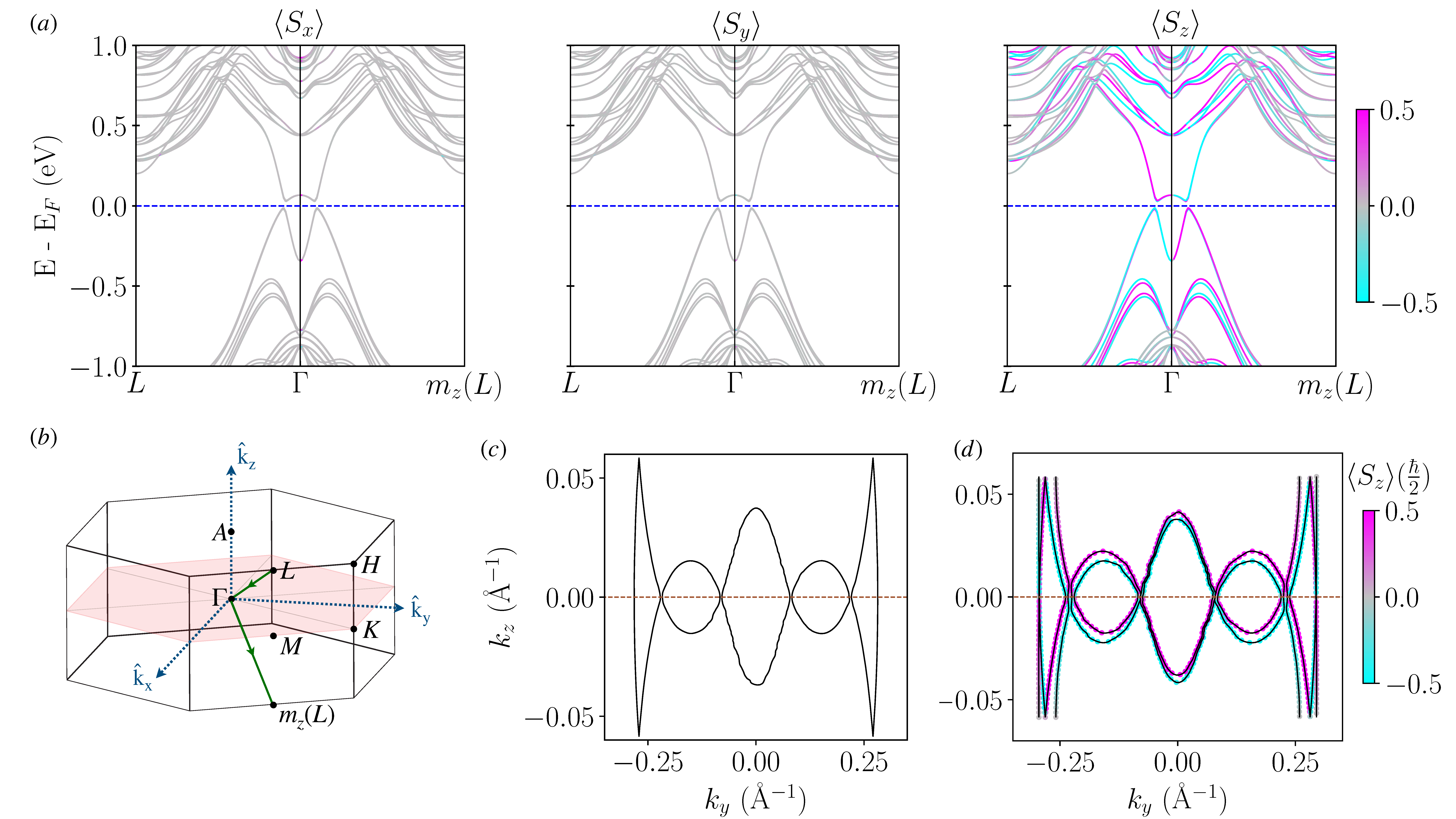}
	\caption{Electronic band structures and momentum-space energy iso-surfaces without SOC of the non-collinear helical \ch{EuIn2As2} phase. (a) Energy bands along the momentum space path $L-\Gamma-m_z(L)$. Spin un-polarized bands for $S_x$ and $S_y$ spin projections, and spin-polarized bands with visible spin splittings for the $S_z$ spin component. (b) The hexagonal first Brillouin zone (1BZ) and the single nodal plane at $k_z=0$ (red) allowed for the $S_z$ spin component. (c) Constant energy cut at $E=E_F-0.60 \ eV$ in the momentum-space $k_x=0$ plane for the non-magnetic crystal and (d) for the helical order. The non-magnetic unit cell has been tripled along the $z-$axis for comparison reasons.}
 \label{helical}
\end{figure*}

In this work, we systematically analyze the spin symmetries of individual helical and broken-helical phases reported for \ch{EuIn2As2}. We uncover several key similarities and notable contrasting characteristics between the two phases. Through our spin symmetry analysis, we identify the odd-parity wave order of the out-of-plane spin polarization in both helical phases, and we identify the g-wave order of the in-plane spin polarization exclusively in the broken-helical phase. These magnetic orders are further confirmed by non-relativistic density functional theory (DFT) calculations and remain mostly unchanged under SOC effects. Finally, we validate the odd-parity nature of these phases by computing the non-equilibrium current-induced spin density -- known as the Edelstein effect. Our results reveal a distinct out-of-plane linear Edelstein response with a predominantly non-relativistic origin. %a phenomenon that has been primarily linked with spin-orbit coupling. 

This article is organized as follows. In Secs. \ref{sectionII} and \ref{sectionIII}, we characterize the non-relativistic spin-momentum locking for the helical and broken-helical phases, employing a spin symmetry analysis and \textit{ab initio} calculations. In Section \ref{sectionIV}, we introduce the linear Edelstein effect in non-centrosymmetric systems, discuss the symmetry-imposed spin density, and validate the findings with computed Edelstein effect for both phases. Finally, in the Section \ref{sectionV} we present our conclusions.

%%%%%%%%%%%%%%%%%%%%%%%%%%%%%%%%%%%            %%%%%%%%%%%%%%%%%%%%%%%%%%%%%%%%%%% 
%%%%%%%%%%%%%%%%%%%%%%%%%%%%%%%%%%% RESULTS   %%%%%%%%%%%%%%%%%%%%%%%%%%%%%%%%%%% 

\begin{figure*}
	\includegraphics[width=0.95\textwidth]{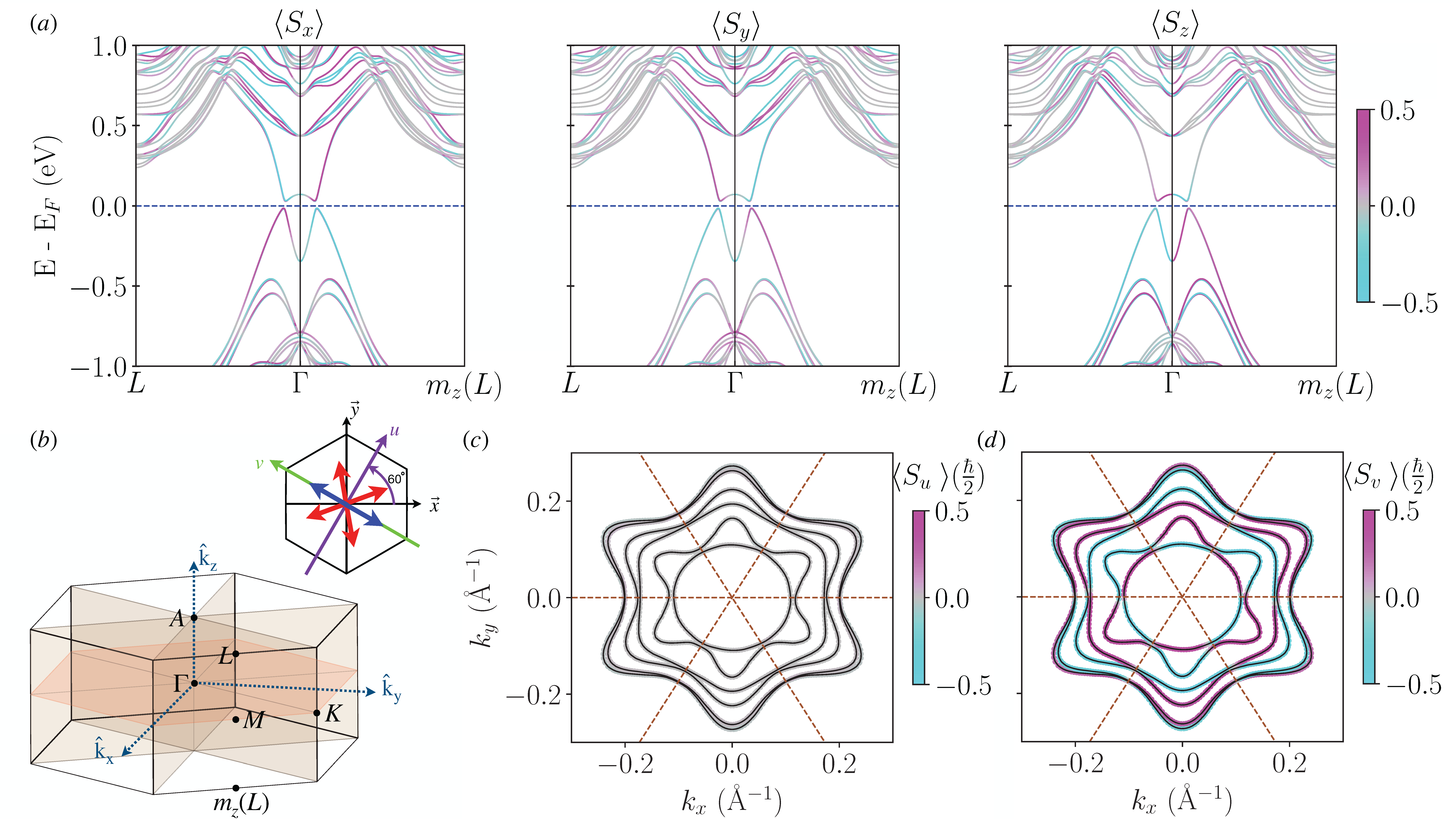}
	\caption{Electronic band structure and energy iso-surfaces without SOC for the non-collinear broken-helical \ch{EuIn2As2} phase. (a) Spin-polarized energy bands along $x$, $y$, and $z$  directions, plotted along the momentum path $L-\Gamma-m_z(L)$. (b) The hexagonal BZ and the four nodal planes associated with the g-wave order, each separating regions of opposite spin polarization. The inset in the top right highlights the orthogonal spin axes $u$(violet), and $v$ (green), which are chosen to visualize the in-plane g-wave order. The red and blue arrows represent the directions of the \ch{Eu} magnetic moments. (c) Spin-polarized energy iso-surface at $E=E_F + 0.60$ eV in the momentum-space $k_z=0.2\mathbf{b}_3$ plane. Here, $\mathbf{b}_3$ corresponds to the out-of-plane reciprocal lattice vector. A spin un-polarized energy iso-surface is observed for the $S_u$ spin component, while in panel (d) a purely g-wave spin texture is seen for the $S_v$ spin component. The mustard-dotted lines indicate the nodal planes. }
 \label{broken}
\end{figure*}

\section{Spin symmetry analysis of the helical phase} 
\label{sectionII}
In the non-relativistic regime, where spin and crystal lattice are decoupled, spin space groups (SSGs) provide a proper description of the magnetic part of the electronic spectra \cite{litvin1974spin,litvin1977spin,brinkman1966theory,xiao2024spin,chen2024enumeration}. %These symmetry transformations come as pairs of operators that act on the spin and real space separately. 
The spin-space group can be decomposed as a direct product between the spin-only group and the non-trivial spin-space group, $\mathbf{r}_{s} \times \mathcal{G}_s$ \cite{litvin1974spin,litvin1977spin}. %This factorization is done in such a way that the $\mathcal{G}_s$ contains only unitary symmetries. 
The spin-only group for coplanar non-collinear magnets corresponds to $\mathbf{r}_{s} = \{[E||E],[C_{2\perp}\mathcal{T}||\mathcal{T}]\}$, where $C_{2\perp}$ represents a $180\degree$ spin-space rotation around the axis perpendicular to the coplanar magnetic moments, $E$ represents the identity operator, and $\mathcal{T}$ represents the time reversal symmetry.

To evaluate the symmetry constraints imposed on the electronic band structure around $\Gamma$, one can ignore the translations and therefore focus on the product $\mathbf{r}_{s} \times \mathbf{R}_s$, with $\mathbf{R}_{s}$ being the non-trivial spin point group. The non-trivial spin point group of the helical phase is

\begin{equation}
     \mathbf{R}_{s}^{H} = {}^{3_z}1{}^{2_z}6/{}^{2_x}m {}^{1}m {}^{2_z}m  \text{.}
\end{equation}
It contains 72 elements \cite{shinohara2024algorithm}, and the name of the group captures the information of the generators $[C_{3z}||E]$, $[C_{2z}||C_{6z}]$, $[C_{2x}||m_z]$, $[E||m_x]$ and $[C_{2z}||m_y]$, where on each pair, the first and second element act on spin and real space respectively. Here, $m_i$ ($C_{n,i}$) represents a mirror ($n$-fold rotation) normal to (around) the axis $i$ ((axes and the corresponding non-trivial spin-space group $\mathcal{G}_s$ are specified in Appendix ~\Ref{ssgs})).

Note that the first symbol ${}^{3_z}1$, representing the generator $[C_{3z}||E]$, is not included in the spin-only group $\mathbf{r}_{s}$, because in the spin space group, this element contains a translational component. However, at the level of the spin point group, this symmetry acts exclusively within the spin-space, enforcing a purely out-of-plane spin polarization in the band structure, i.e., $\langle S_x \rangle =0$ and $\langle S_y \rangle =0$. Additionally, the spin point group includes the symmetry $[C_{2z} || C_{2z}]$, which in combination  with the coplanar spin-only group $[C_{2z}\mathcal{T}||\mathcal{T}]$ symmetry, results in $S_z(k_x,k_y,k_z) = -S_z(k_x,k_y,-k_z)$, an antisymmetric $S_z$ spin polarization with a single spin-un-polarized plane at $k_z=0$. 

By employing the spin point group, we can construct a simple two-band Hamiltonian that characterizes the energy dispersion and spin polarization of the helical order around $\Gamma$, expressed as

\begin{equation}
    h(\mathbf{k}) =  \Bigl(A \ (k_x^2 + k_y^2) + B \ k_z^2 \Bigr)\sigma_0 + C \  k_z \sigma_z \,,
    \label{mini_heli}
\end{equation}
where $\mathbf{k}$ represents the wave-vector and $\sigma_z$ is the Pauli matrix along the [001] direction. It is important to note that the antisymmetric $S_z$ spin polarization is fully captured by the odd-parity term, $k_z\sigma_z$, the reason for which it can be referred to as the odd-parity wave order. The linear energy dispersion near $\Gamma$ can be observed if we plot the energy bands along the $A-\Gamma-A$ path, see Appendix ~\Ref{linear}.

We next perform spin-polarized density functional theory (DFT) band structure calculations to confirm the previous findings. All the methodology used here and in the following sections is presented in Appendix ~\Ref{methods}. We select the $L-\Gamma-m_z(L)$ path, which intersects the nodal plane $k_z=0$, as illustrated in Fig. \ref{helical}(b). The energy bands without SOC, shown for each spin projection in Fig. \ref{helical}(a), reveal spin-polarized bands exclusively for the $S_z$ spin component. Since the  electronic structure calculations were performed using the Vienna ab-initio simulation package (VASP), which does not determine the spin symmetries and considers only the unitary part of the magnetic point group, symmetrization was disabled in our DFT calculations. As a result, the negligible small values of the in-plane spin polarization, $S_x$ and $S_y$, observed near the $\Gamma$ point constitute numerical artifacts. The antisymmetric $S_z$ spin polarization can be further visualized in the energy iso-surface along the $k_x=0$ plane, shown in Fig. \ref{helical}(d). A comparison between the energy iso-surfaces of the non-magnetic order in Fig \ref{helical}(c) and the one under helical order reveals that the spin splittings are of non-relativistic origin, arising primarily due to magnetic exchange between local and itinerant electrons. 

We point out that the mechanism responsible for generating an antisymmetric $S_z$ spin polarization in the helical phase differs from the materials studied in \cite{hellenes2024pwavemagnets}. The main distinction is that those materials preserve $\mathcal{T}\tau$, which enforces an antisymmetric spin polarization for all spin components, and this magnetic order is preserved even when SOC effects are included. In contrast, in the helical phase, where $\mathcal{T}\tau$ is broken, only the coplanar symmetry $[C_{2z} \mathcal{T}||\mathcal{T}]$ guarantees an antisymmetric spin polarization for the $S_z$ component, while the in-plane spin components exhibit a symmetric spin polarization. However, there is an additional symmetry $[C_{3z}||E]$ that forces the in-plane components to vanish, resulting in an effective antisymmetric and collinear spin polarization in momentum space, but only in the limit of zero SOC.

\section{Spin symmetry analysis of the broken-helical phase} 
\label{sectionIII}
This more complex magnetic configuration, results in a smaller non-trivial spin space group, see App.~\Ref{ssgs}. Excluding the translations, the non-trivial spin point group is:

\begin{equation}
    \mathbf{R}_{s}^{B} = {}^{2_z}6/{}^{2_u}m {}^{1}m {}^{2_z}m \text{.}
    \label{spg_br0}
\end{equation}
It has 24 elements \cite{shinohara2024algorithm}, and it can be recognized as the spin point group No. 478 of Table 1. in Ref. \cite{litvin1974spin}. It has a structure similar to that of the group of the helical phase, but now the ${}^{3_z}1$ symbol is not present, because the 3-fold spin rotation is broken. The generators are $[C_{2z}||C_{6z}]$, $[C_{2x}||m_z]$, $[E||m_u]$ and $[C_{2z}||m_v]$, with $u$ and $v$ being the in-plane orthogonal directions shown in Figure \ref{broken}(b) in the top-right side. This group can be rewritten using conventional crystallographic point groups, in the following form:

\begin{equation}
    \mathbf{R}_{s}^{B} =  \Bigl( [E|| 3m] +  [C_{2z} || 6mm-3m] \Bigr) \times \mathbb{Z}_2^{[C_{2v}||\mathcal{P}]} \,,
    \label{spg_br}
\end{equation}
where $6mm$ and $3m$ are crystallographic point groups defined in  Appendix ~\Ref{ssgs}, and $6mm-3m$ is the set difference. The term inside the parentheses is itself a group, and we do the direct product with the binary group $\mathbb{Z}_2^{[C_{2v}||\mathcal{P}]} = \{[E||E],[C_{2v}||\mathcal{P}]\}$, where $\mathcal{P}$ denotes the inversion operator and $C_{2v}$ a twofold rotation around the $v$-axis.

Focusing on $\mathbf{R}_{s}^{B}$, we can identify a similarity between the first term in parentheses and the spin point group of the collinear g-wave altermagnetic phase of \ch{EuIn2As2}, see Appendix ~\Ref{coll}. The spin rotation $C_{2z}$ has an axis orthogonal to the in-plane spins, analogous to the $C_2$ spin rotation in an altermagnetic point group \cite{vsmejkal2022beyond}. This suggests that the in-plane spin components can exhibit a $g$-wave order. The symmetry $[C_{2v}||\mathcal{P}]$, which enforces an antisymmetric spin polarization for $S_u$, in combination with the symmetric order imposed by the g-wave order, will fully supress the spin polarization for $S_u$ while keeping invariant a g-wave order for $S_v$. Regardless of the choice of the spin axis, the four spin-degenerate nodal planes of the g-wave order are protected by the following transposing mirror symmetries: $[C_{2z}||m_{v}]$, $[C_{2z}||m_{v'}]$,$[C_{2z}||m_{v''}]$, and $[C_{2z}||m_{z}]$. %The axes $v'$, and $v''$, are defined in App.~\Ref{ssgs}. 
Finally, similar to the helical order,  the symmetry $[C_{2z} || C_{2z}]$ is preserved, indicating that an odd-parity wave magnetic order should be expected for the spin component $S_z$, with a single nodal plane at $k_z=0$.

A minimal two-band model capturing the spin symmetries of the broken helical is given by the following Hamiltonian:

\begin{multline}
    h(\mathbf{k}) = \Bigl(A \ (k_x^2 + k_y^2) + B \  k_z^2 \Bigr)\sigma_0  \ + \ C \ k_z \sigma_z \\
   \  +  \ D \  k_yk_z\Bigl( (\sqrt{3}k_x)^2 -k_y^2\Bigr) \ \vec{\mathbf{\sigma}}\cdot\hat{v} \,,
\end{multline}
where $\hat{v} = \{1,-1/\sqrt{3},0\}$ is an in-plane vector that points along the $v$ axis and $\sigma_0$ represents the identity $2\times2$ matrix. In contrast to the helical phase, the broken helical introduces a third term in the effective model, which captures the g-wave order of the $S_v$ spin polarization. The emergence of an even-parity order is expected, as $\mathcal{T}\tau$ is broken.

In Fig. \ref{broken}, we present the DFT calculations for the broken-helical phase. The 1BZ with four nodal planes corresponding to the g-wave order is illustrated in Fig. \ref{broken}(b). To visualize the alternating spin polarization in the energy bands, we select the path $L-\Gamma-m_z(L)$. In Fig. \ref{broken}(a), we plot the spin-polarized bands setting the spin-axis along the conventional $xyz$ axis. Unlike the helical phase, we now observe in-plane spin splittings, while the antisymmetric spin splittings for the $S_z$ spin component, are significantly reduced. Finally, by aligning the spin-axis along the $uv$ axis, we calculate the energy iso-surfaces shown in Fig \ref{broken}(c,d), which reveal un-polarized bands for $S_{u}$ and a pure g-wave order for $S_{v}$, consistent with our spin symmetry analysis.

\section{Non-relativistic linear Edelstein effect}
\label{sectionIV}
The Edelstein effect refers to the emergence of a non-equilibrium spin density induced by an applied electric field in non-centrosymmetric systems. Traditionally, this effect has been associated with spin-orbit coupling \cite{manchon2019current,bihlmayer2022rashba,vzelezny2014relativistic,leiva2023spin}. However, recent studies have demonstrated that a non-relativistic Edelstein effect can also occur in non-collinear non-centrosymmetric compounds with an antisymmetric spin texture or coplanar p-wave magnets \cite{gonzalez2024non, chakraborty2024highly}.

The non-equilibrium spin density can be calculated using the Kubo linear response formalism, \cite{garate2009influence, manchon2019current}. %Here, we consider terms only up to linear order in the electric field,
i.e., $\delta s_i = \chi_{ij}E_{j}$ with $\chi_{ij}$ being the response tensor and $E_j$ the electric field. The response tensor can be decomposed into two components, the $\mathcal{T}$-even (intraband) and $\mathcal{T}$-odd (interband) contributions \cite{manchon2019current,li2015intraband}. The intraband term is given by 

\begin{equation}
\chi^{intra}_{ij} = \frac{e\hbar}{2\Gamma} \int \frac{d\mathbf{k}}{(2\pi)^3} \sum_{n} \mathrm{Re}[( \hat{s}_i)_{nn}(\hat{v}_{j})_{nn}] 
\times \delta({E_{\mathbf{k}n}-E_F})\,,
\end{equation}
with $e$ being the electric charge, $E_{\mathbf{k}n}$ the n-th energy band for a given $\mathbf{k}$ wave vector, $E_F$ the Fermi energy, $\hat{s}_i$ the spin operator, $\hat{v}_j$ the velocity operator, and $\Gamma$ the spectral broadening due to disorder.

The interband term is given by

\begin{multline}
\chi^{inter}_{ij} = 2e\hbar \int \frac{d\mathbf{k}}{(2\pi)^3} \sum_{n\neq m} (f_{\mathbf{k}n}-f_{\mathbf{k}m}) \mathrm{Im}[(s_i)_{nm}( v_{j})_{mn}] \\
\times \frac{(E_{\mathbf{k}n}-E_{\mathbf{k}m})^2 -\Gamma^2}{[(E_{\mathbf{k}n}-E_{\mathbf{k}m})^2+\Gamma^2]^2}\,,
\end{multline}
with $f_{\mathbf{k}n}$ being the Fermi distribution function, and $n$ and $m$ correspond to different bands. 

As previously reported, the helical and broken-helical orders of \ch{EuIn2As2} exhibit a non-relativistic antisymmetric spin texture for the $S_z$ spin component while simultaneously breaking $\mathcal{P}$ symmetry. As a result, a non-relativistic Edelstein effect is expected in both helical phases. To begin, we analyze the symmetry imposed shape of the response tensor $\chi_{ij}$ under the spin-symmetry generators (see Table \ref{table}). In the absence of spin-orbit coupling, only the intraband contribution along the lattice vector $\mathbf{c}$, $\chi^{intra}_{cc}$, is allowed for both non-collinear orders. This suggests that an electric field applied along the $z$ direction can generate a non-equilibrium out-of-plane spin population. When spin-orbit coupling is introduced, additional in-plane intraband components are allowed, and the interband contribution becomes possible. 

\begin{table} %[ht]
\centering
        \begin{tabular}{cccc}
            \hline
            \hline
            & & Without SOC & SOC \\
            \hline
            Helical& $\chi^{intra}=$ &  $\begin{pmatrix} 0 & 0 & 0\\0 & 0 &0 \\0&0&\chi_{c}\end{pmatrix}$ & $\begin{pmatrix} \chi_{a} & \chi_a/2 & 0\\ \chi_a/2 & \chi_a &0 \\0&0&\chi_{c}\end{pmatrix}$ \\ \\
            & $\chi^{inter}=$ &  $\begin{pmatrix} 0 & 0 & 0\\0 & 0 &0 \\0&0&0\end{pmatrix}$ & $\begin{pmatrix}0 & \chi_{b} & 0\\ -\chi_{b} & 0 &0 \\0&0&0\end{pmatrix}$ \\  \\\hline
         Broken-Helical & $\chi^{intra}=$ &  $\begin{pmatrix} 0 & 0 & 0\\0 & 0 &0 \\0&0&\tilde{\chi}_{c}\end{pmatrix}$ & $\begin{pmatrix} \tilde{\chi}_{a} & \tilde{\chi}_b & 0\\ \tilde{\chi}_b & \tilde{\chi}_{a} &0 \\0&0&\tilde{\chi}_{c}\end{pmatrix}$ \\   \\
& $\chi^{inter}=$ &  $\begin{pmatrix} 0 & 0 & 0\\0 & 0 &0 \\0&0&0\end{pmatrix}$ & $\begin{pmatrix} \tilde{\chi}_a& \tilde{\chi}_{b}& 0\\  -\tilde{\chi}_{b}& -\tilde{\chi}_a  &0 \\0&0&0\end{pmatrix}$ \\\hline
        \end{tabular}
\caption{Shape of the intraband and interband response tensor $\chi$ for the helical and broken-helical orders expressed as a function of the lattice vectors, $\mathbf{a}=a\mathbf{\hat{x}}$, $\mathbf{b}=-a\frac{\sqrt{3}}{2}\mathbf{\hat{x}} +\frac{a}{2}\mathbf{\hat{y}}$, and $\mathbf{c}=c~\mathbf{\hat{z}}$. In the non-relativistic case, the spin-point group generators are applied while for the relativistic case, magnetic-point group generators are employed.} 
\label{table}
\end{table}
% \vspace*{2cm}

\begin{figure}[h]
	\includegraphics[width=0.4\textwidth]{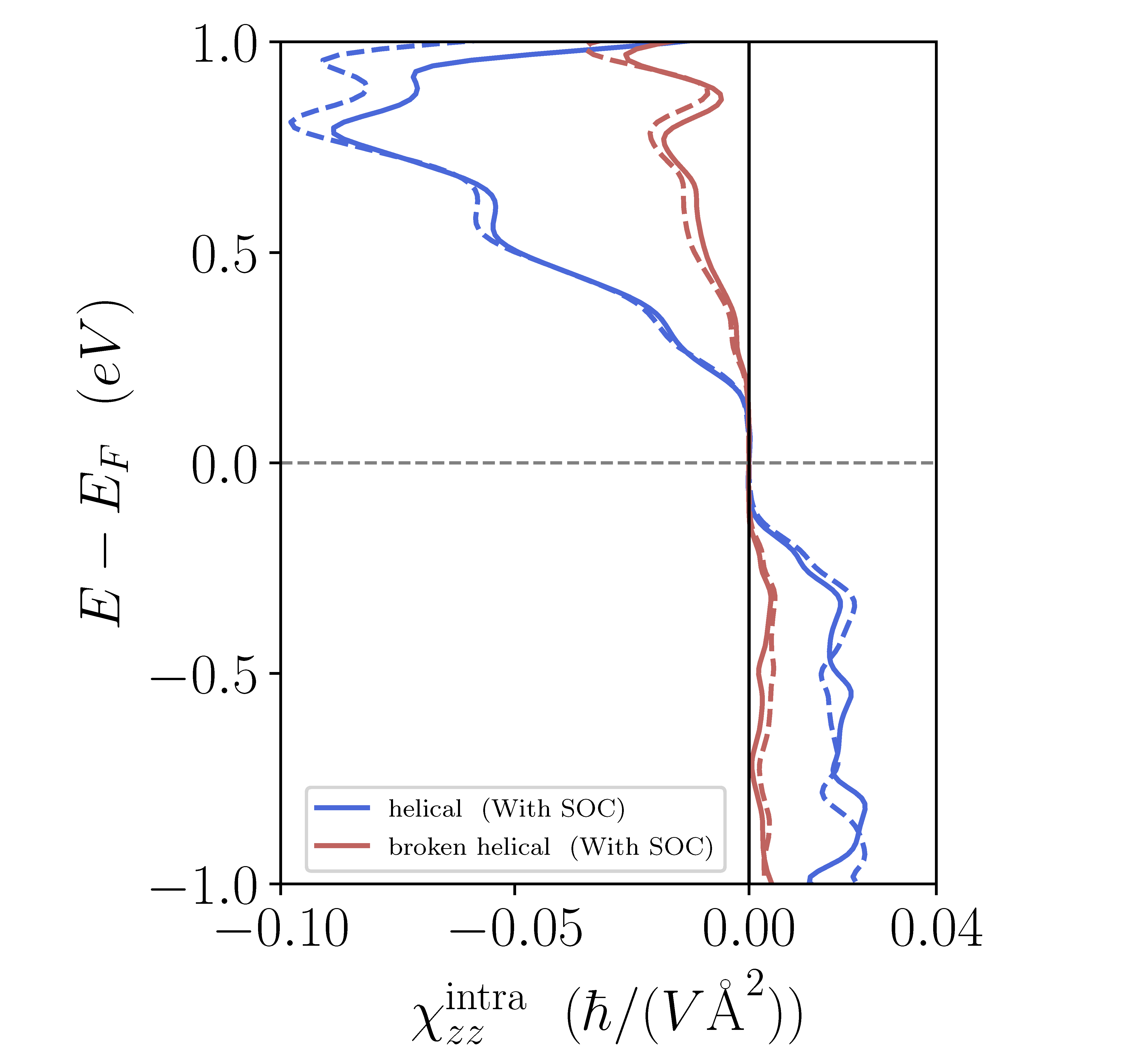}
	\caption{Computed intraband out-of-plane response tensor $\chi_{zz}^{intra}$. Calculations with (solid line) and without (dashed line) spin-orbit coupling of both the helical and broken-helical phases considering a disorder $\Gamma = 0.01$ eV.}
 \label{intraband}
\end{figure}

From the symmetry analysis, we find that only the intraband contribution is directly linked to the non-relativistic antisymmetric spin-momentum locking. We report this contribution with and without SOC in Fig. \ref{intraband}. Within the energy window $E=E_F \pm 1 \ eV$, the helical phase exhibits spin density $\delta s_z$ values larger than those in the broken-helical phase, which is consistent with the observation of $S_z$ spin splittings in the helical phase with magnitude larger than those in the broken-helical. Near the Fermi level, $E=E_F\pm 0.3 \ eV$, the magnitude of the linear intraband responses can be approximately related by a factor of $\chi_{zz}^{\mathrm{helical}}/\chi_{zz}^{\mathrm{broken}} \approx 5$. Introducing spin-orbit coupling has little impact on the out-of-plane spin density values, which aligns with the subtle breaking of the antisymmetric $S_z$ spin texture upon including SOC, see Appendix ~\Ref{soc_effects}. Nevertheless, SOC gives rise to in-plane spin density terms, which are approximately an order of magnitude smaller than the dominant out-of-plane response, see Appendix ~\Ref{ed_soc}. The significant difference in the out-of-plane polarized spin density values between both phases could serve as a signature for detecting the magnetic transition via transport measurements, in contrast to other transport quantities, such as the longitudinal conductivity, which also shows differences between both phases but to a lesser extent, see Appendix ~\Ref{cond}. Although the spin density vanishes at the Fermi level, \ch{EuIn2As2} is experimentally known to be highly hole-doped \cite{yan2022field,regmi2020temperature,yan2024doping}, which could already lead to a sizable Edelstein response.
%While the spin density vanishes at the Fermi level, being a semiconductor material, doping can shift the Fermi level \cite{yan2024doping} and induce sizeable values of the Edelstein reponse.

A final important point with regards to the measurement of the Edelstein effect, is that its presence -- or lack of presence -- will also serve to validate or discard the proposed amplitude modulated phases in Ref. \cite{donoway2024multimodal}, since both their A$_1$ ($\mathcal{P}$ symmetric) and A$_2$ ($\mathcal{PT}$ symmetric) phases would not have an Edelstein effect.

\section{Conclusion}
\label{sectionV}

We studied two non-collinear coplanar magnetic configurations reported in \ch{EuIn2As2}: the helical and broken-helical phases. For each phase, we conducted a spin symmetry analysis and validated our theoretical predictions with non-relativistic first-principles calculations, which showed good agreement between the two approaches. Our findings highlight the potential of spin symmetries to capture the exchange-dominated physics and to characterize unconventional magnetic phases in systems with non-collinear spin arrangements, even in the presence of strong SOC.

We identified an out-of-plane odd-parity wave order, featuring a single nodal plane in momentum-space, present in both helical phases. Additionally, we report an in-plane g-wave magnetic order, but only in the broken-helical phase, which opens new avenues for investigating the interplay between the g-wave order parameter and the topology of the axion insulator state in \ch{EuIn2As2}. Notably, these odd- and even-parity magnetic orders remain mostly unchanged in the presence of SOC, suggesting they could be potentially probed through spectroscopic techniques.

%Finally, motivated by identifying another distinguishing feature 
To further differentiate between the two helical phases and to validate the odd-parity nature of the magnetic order, we calculated the linear Edelstein effect, a response allowed by symmetries in both configurations. Our results revealed a distinct out-of-plane current-induced spin density with a predominantly non-relativistic origin. Remarkably, this anisotropic response exhibits a large difference in magnitude between the two helical phases, which could serve as an alternative indicator of the magnetic phase transition.

%It may also help differentiate them between the magnetic ground states recently proposed by Donoway et al. \cite{donoway2024multimodal} involving amplitude modulation (phase A$_1$ and A$_2$), which do not exhibit a linear Edelstein response, even in the presence of spin-orbit coupling.

\textbf{Acknowledgments---} This work is supported by the Deutsche Forschungsgemeinschaft (DFG, German Research Foundation) — TRR 288 – 422213477 (project B05 and A09). A.C. acknowledges financial support from Alexander von Humboldt foundation. L.\v{S}. acknowledges support from the ERC Starting Grant No. 101165122. The authors gratefully acknowledge the computing time granted on the supercomputer MOGON 2 at Johannes Gutenberg University Mainz (hpc.uni-mainz.de). We acknowledge fruitful discussions with Rafael González-Hernández, Veronika Sunko and Anna Birk Hellenes.

\appendix

\section{Computational Methods}
\label{methods}
%\textit{Computational methods ---} 
The electronic band structures were performed within DFT, using the Vienna ab-initio Simulation Package (VASP) \cite{kresse1996efficient, kresse1999ultrasoft}, and the Perdew-Burke-Ernzerhof (PBE) exchange-correlation functional was applied as the generalized gradient approximation (GGA) \cite{perdew1996generalized}. To capture the non-relativistic effects, both the spin-orbit coupling and the symmetrization were switched off  by setting $\mathrm{LSORBIT=FALSE}$ and $\mathrm{ISYM}=-1$ in VASP. In both scenarios, with and without SOC, we allow for the noncollinearity by setting $\mathrm{LNONCOLLINEAR = TRUE}$. A Hubbard $U$ of 5 eV was included to account for the localized nature of the Eu 4f orbitals. The Brillouin zone was sampled with a $\Gamma$-centered k-point mesh of $13\times13\times3$, along with a Gaussian smearing of 0.05 eV, a kinetic energy cutoff of 250 eV, and a convergence criterion of $10^{-6}$ eV were used. The self-consistent calculations were carried out using constraints in the magnetic moment's directions as reported in MAGNDATA\cite{gallego2016magndata,magndata_data}. Applying constraints introduces a penalty energy, which indicates the deviation from the desired magnetic moments \cite{ma2015constrained}. The penalty energies achieved for both the helical and broken-helical orders were $10^{-8}$ eV. 

The spin-polarized energy bands and energy iso-surface plots were generated using PyProcar \cite{herath2020pyprocar}. Additionally, we employed the \textit{spinspg} package \cite{shinohara2024algorithm} to determine the corresponding spin symmetries. For further data analysis and post-processing, a tight-binding Hamiltonian was generated within the projected Wannier functions for Eu-$d,f$, In-$s,p$, and As-$p$ orbitals using the Wannier90 package \cite{pizzi2020wannier90}. The charge-induced spin density was then calculated using the Wannierberri package \cite{tsirkin2021high} with a finite momentum k-mesh of $162\times162\times12$.

\section{Spin space groups}
\label{ssgs}

\begin{figure}[H]
\centering
	\includegraphics[width=0.3\textwidth]{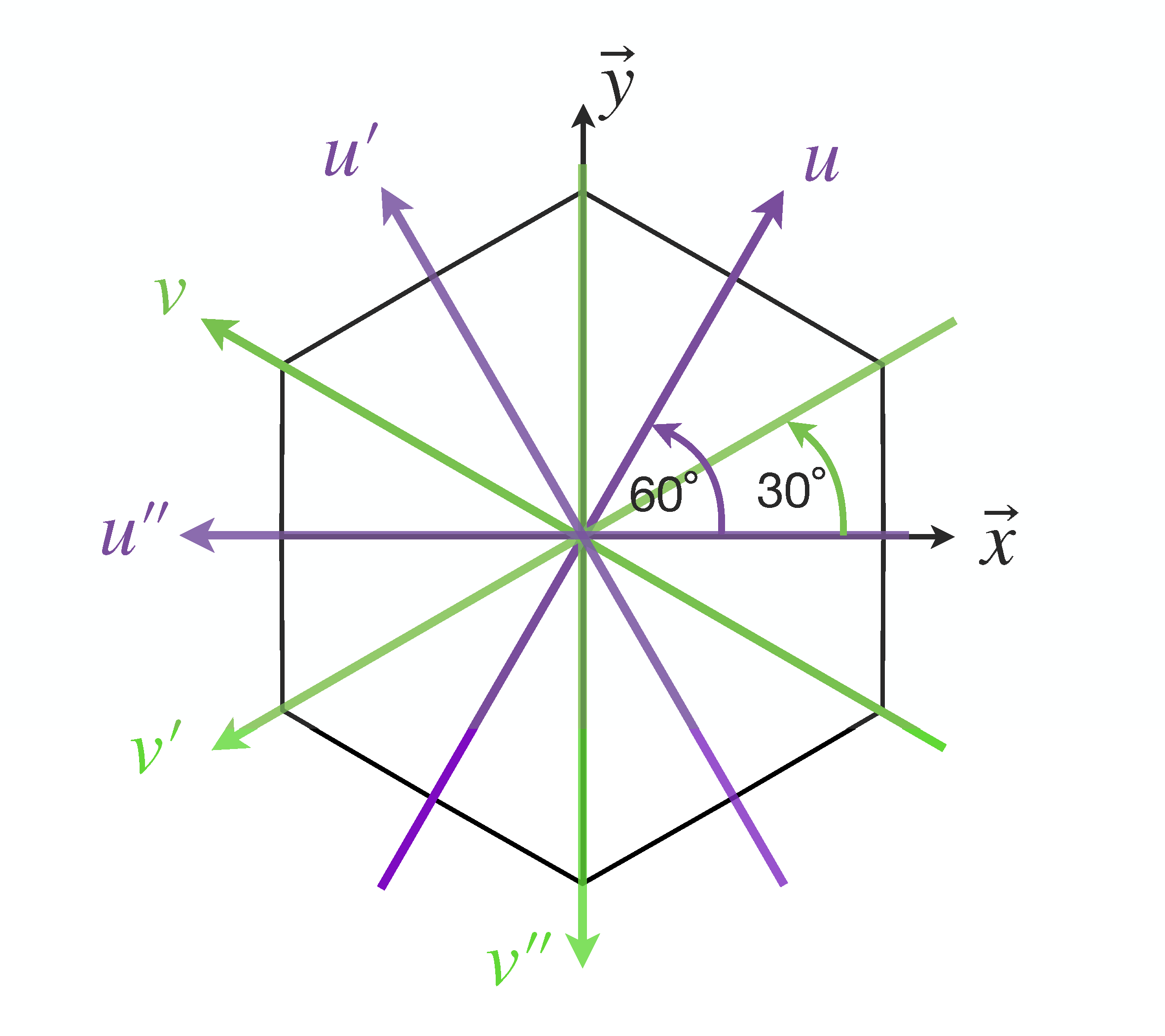}
	\caption{Six in-plane axes, $\{u,u',u''\}$ (green), and  $\{v,v',v''\}$ (violet), where each pair $\{u,v\}$, $\{u',v'\}$, and $\{u'',v''\}$ consists of orthogonal axes.}
 \label{axes}
\end{figure}

A spin space group can be decomposed as $\mathbf{r}_s\times \mathcal{G}_S$ \cite{litvin1974spin,litvin1977spin,brinkman1966theory,xiao2024spin,chen2024enumeration}, with $\mathbf{r}_s$ denoting the spin-only group, and $\mathcal{G}_S$ the nontrival spin space group. Since the magnetic orders studied in this work are coplanar, the spin-only group will correspond to $\mathbf{r}_{s}=\{[E||E], [C_{2z}\mathcal{T}|\mathcal{T}]\}$.
A general element of the nontrivial spin space group can be written as $[R_s||R_r| \tau]$, where $R_s$ and $R_r$ are point symmetries acting on spin and real space respectively, and $\tau$ is a translation. After factorizing the group structure by the taking out the Bravais lattice translations, the non trivial spin space group of the helical order can be written as 

\begin{multline*}
\mathcal{G}_{s}^{H} = \Bigl( \  [E||\mathbf{H}|0] \ + \ [C_{2v}||\mathcal{P} \cdot \mathbf{H}|0]  \ + [C_{6z}||\mathbf{G}-\mathbf{H}|\tau] \ +   \\[6pt] 
[C_{2u''}||\mathcal{P} \cdot (\mathbf{G}-\mathbf{H})|\tau] \ + [C_{3z}||\mathbf{H}|2\tau] \ + \ [C_{2v'}||\mathcal{P} \cdot \mathbf{H}|2\tau] \  +  \\[6pt] 
[C_{2z}||\mathbf{G}-\mathbf{H}|3\tau] \ + \ [C_{2u}||\mathcal{P} \cdot (\mathbf{G} -\mathbf{H})|3\tau] \ + \\[6pt]
[C_{3z}^{-1} || \mathbf{H}|4\tau] \ + \ [C_{2v''}||\mathcal{P} \cdot \mathbf{H}|4\tau]  \ +  \\[6pt] [C_{6z}^{-1}||\mathbf{G}-\mathbf{H}|5\tau] \ + \ [C_{2u'}||\mathcal{P} \cdot (\mathbf{G}-\mathbf{H})|5\tau] \ \Bigr)\,,
\end{multline*}
where $\mathbf{G}=6mm = \{E, C_{3z}, C_{3z}^{-1}, C_{6z},C_{6z}^{-1},C_{2z},m_{u}, m_{u'},$
$ m_{u''},m_{v}, m_{v'}, m_{v''}\}$ and $\mathbf{H} =3m =\{ E,C_{3z},C_{3z}^{-1},m_{u},m_{u'},$
$m_{u''}\}$ are the crystallographic point groups acting on real space, where $E$ denotes the identity operator, $C_{nz}$ a n-fold rotation around the z-axis, and $m_{\hat{\mathbf{n}}}$ a mirror operator with respect to a plane normal to the unit vector $\hat{\mathbf{n}}$ defined in Fig. \ref{axes}. In the expression of $\mathcal{G}_{s}^{H}$, we organize each of the symmetries acting on spin space together with corresponding set of symmetries acting on real space. The real-space symmetry sets involve the already mentioned group $\mathbf{G}$, its halving subgroup $\mathbf{H}$, and the complement set $\mathbf{G}-\mathbf{H} = \{ C_{6z},C_{6z}^{-1},C_{2z}, m_{v}, m_{v'}, m_{v''}\}$, which consists of elements that are contained in $\mathbf{G}$ but not in $\mathbf{H}$. In addition, we also include the sets that result after applying inversion $\mathcal{P}$ symmetry to the real-space sets.
For the spin-space part of the symmetries, $C_{n \hat{\mathbf{n}}}$ denotes an $n$-fold rotation around the $\hat{\mathbf{n}}$-axis defined in Fig. \ref{axes}. All the translations involved are multiples of $\tau=(0,0,\frac{1}{6}c)$ with $c$ being the Brillouin zone length along $\hat{z}$. The generators of the helical spin space group are $[C_{6z}||C_{6z}| \tau], [C_{6z}||m_{v}|\tau], [C_{2v}||\mathcal{P}]$, and $[C_{2z}\mathcal{T}||\mathcal{T}]$. Here, the last generator is anti-unitary (it involves time-reversal $\mathcal{T}$), and it comes from the spin-only group.

Analogously, the non-trivial spin-space group (modulo the Bravais lattice translations) of the broken-helical order can be written as
\begin{multline*}
    \mathcal{G}_{s}^B= \Bigl( \ [E || \mathbf{H}] \ +
    [C_{2z}|| (\mathbf{G}-\mathbf{H})|3\tau] \  + \\[6pt]    
    [C_{2v}|| \mathcal{P} \cdot \mathbf{H}| 4\tau] \ +
    [C_{2u} || \mathcal{P} \cdot (\mathbf{G} -\mathbf{H}) | \tau] \Bigr)\,.
\end{multline*}
Here, when the translations are ignored, one can rearrange the elements acting on spin space, to recover the expression used for the spin point group in Eq. \ref{spg_br}. The broken-helical spin space group generators are given by the symmetries $ [C_{2z}||C_{6z}| 3\tau],[C_{2z}||m_{v}|3\tau],[C_{2v}||P|4\tau]$, and $[C_{2z}\mathcal{T}|| \mathcal{T}] $.

\section{Linear energy dispersion}
\begin{figure}[h]
\centering
	\includegraphics[width=0.5\textwidth]{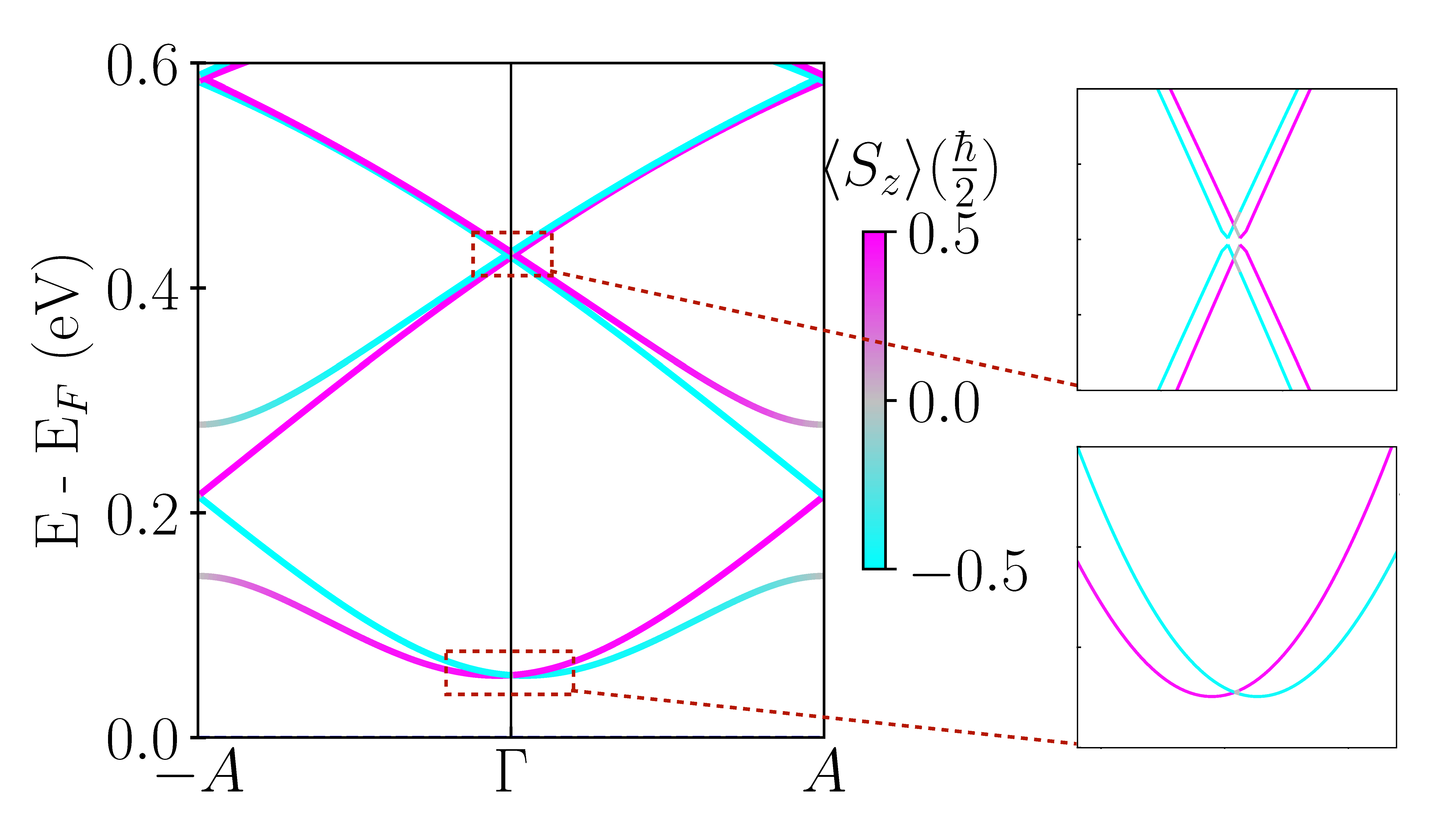}
	\caption{$S_z$ spin polarized energy bands without spin-orbit coupling of the helical phase along the  $A-\Gamma-A$ path. The small plots on the right are zoomed-in views near $\Gamma$. }
 \label{linear}
\end{figure}

The linear energy dispersion near $\Gamma$, arising from the linear term $k_z\sigma_z$ in Eq. \ref{mini_heli}, can be visualized along the $A-\Gamma-A$ path, as shown in Fig. \ref{linear}.

\section{Collinear phase of \ch{EuIn2As2} }
\label{coll}
\begin{figure}[h]
\centering
	\includegraphics[width=0.5\textwidth]{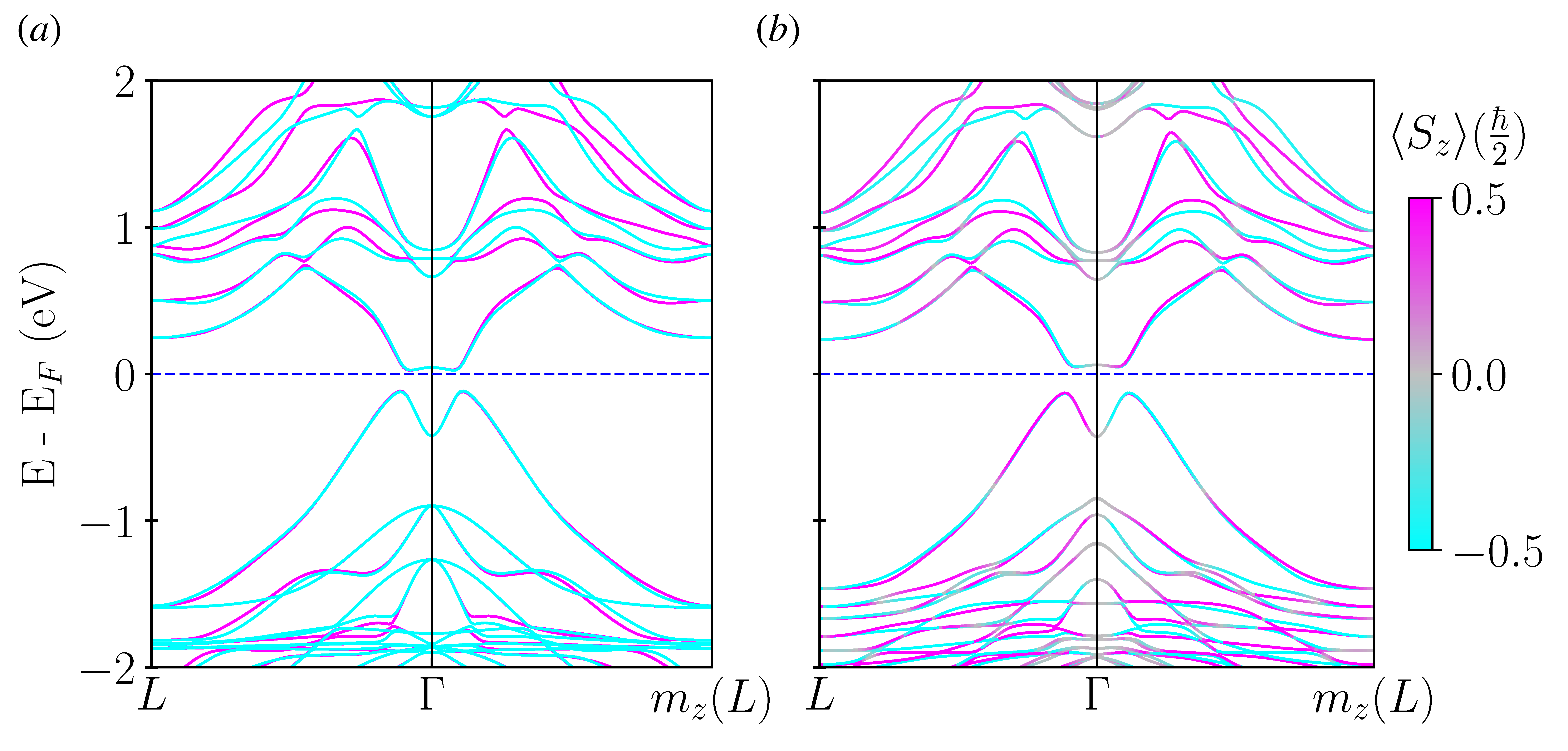}
	\caption{Electronic band structure of the collinear \ch{EuIn2As2} order along $L-\Gamma-m_z(L)$ path, (a) without and (b) with spin-orbit coupling. The spin splittings are of non-relativistic origin, with values reaching up to 100 meV.}
 \label{collinear}
\end{figure}

The collinear magnetic order of \ch{EuIn2As2} is described by the nontrivial spin Laue group $\mathbf{R}_S = [E|| \bar{3}m] + [C_2||6/mmm -\bar{3}m]$, indicating a bulk g-wave altermagnetic order that enables four spin un-polarized nodal planes. The opposite spin sub-latices are connected by the spin symmetries $[C_{2}||m_v]$,$[C_{2}||m_{v'}]$, $[C_{2}||m_{v''}]$, and $[C_{2}||m_z]$. In Figure \ref{collinear}, we show the electronic band structures of the collinear magnetic order with and without SOC, revealing that the largest spin splittings are primarily of non-relativistic origin and can reach magnitudes up to 100 meV.

\section{Spin-ordering under SOC effects}
\label{soc_effects}
As spin-orbit coupling plays a significant role in \ch{EuIn2As2} due to its strongly localized f-orbitals, it is necessary to analyze the effects of including SOC on the non-relativistic odd-wave and g-wave spin-orderings together with their spin splittings. .
\begin{figure}[h]
\centering
	\includegraphics[width=0.5\textwidth]{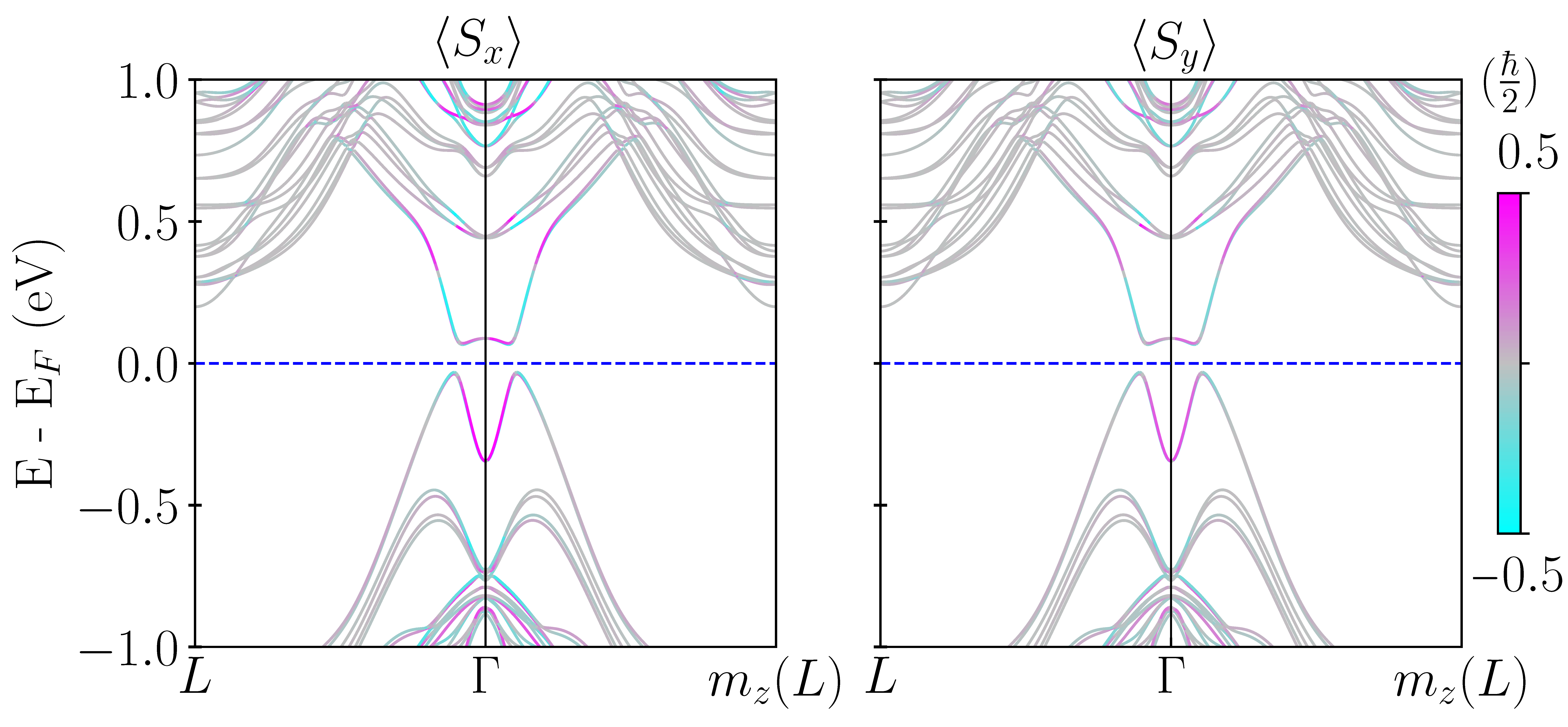}
	\caption{In-plane spin polarized energy bands of the helical phase along the $L-\Gamma-m_z(L)$ path including spin-orbit coupling.}
 \label{soc_helical}
\end{figure}

\begin{figure}[h]
\centering
	\includegraphics[width=0.5\textwidth]{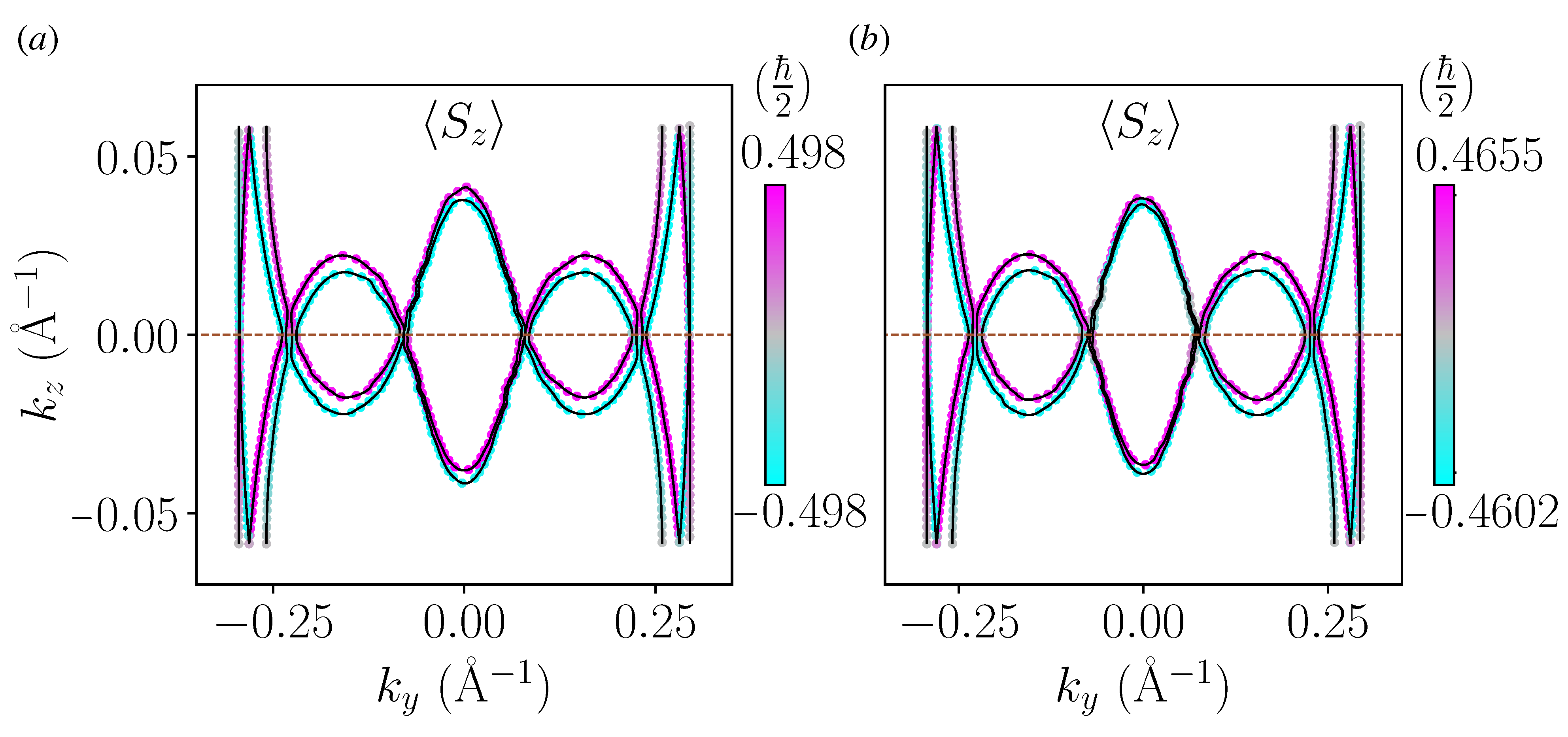}
	\caption{Momentum-space $S_z$ spin polarized energy iso-surfaces of the helical phase in the $k_x=0$ plane for an energy $E=E_F-0.60$ eV without SOC (left) and with SOC (right). In the non-relativistic regime, the antisymmetric spin polarization is protected by the coplanar symmetry $[C_{2z}\mathcal{T}||\mathcal{T}]$.}
 \label{odd}
\end{figure}
When SOC is included, the spin symmetries that were initially protecting the non-relativistic magnetic orders are broken. In the helical phase, the breaking of $[C_{3z} ||E]$, which in the non-relativistic regime supressed the in-plane spin polarization, now allows it, as shown in Fig. \ref{soc_helical}. Regarding the $S_z$ spin polarization, in Fig. \ref{odd} we show a comparison of the energy iso-surfaces with and without SOC. At first glance, both energy iso-surfaces appear very identical; however, a closer inspection of the minimum and maximum values of $S_z$ reveal a subtle breaking of the antisymmetric spin texture in the third decimal place, which results from the breaking of the coplanar symmetry. 

\begin{figure}[H]
\centering
	\includegraphics[width=0.5\textwidth]{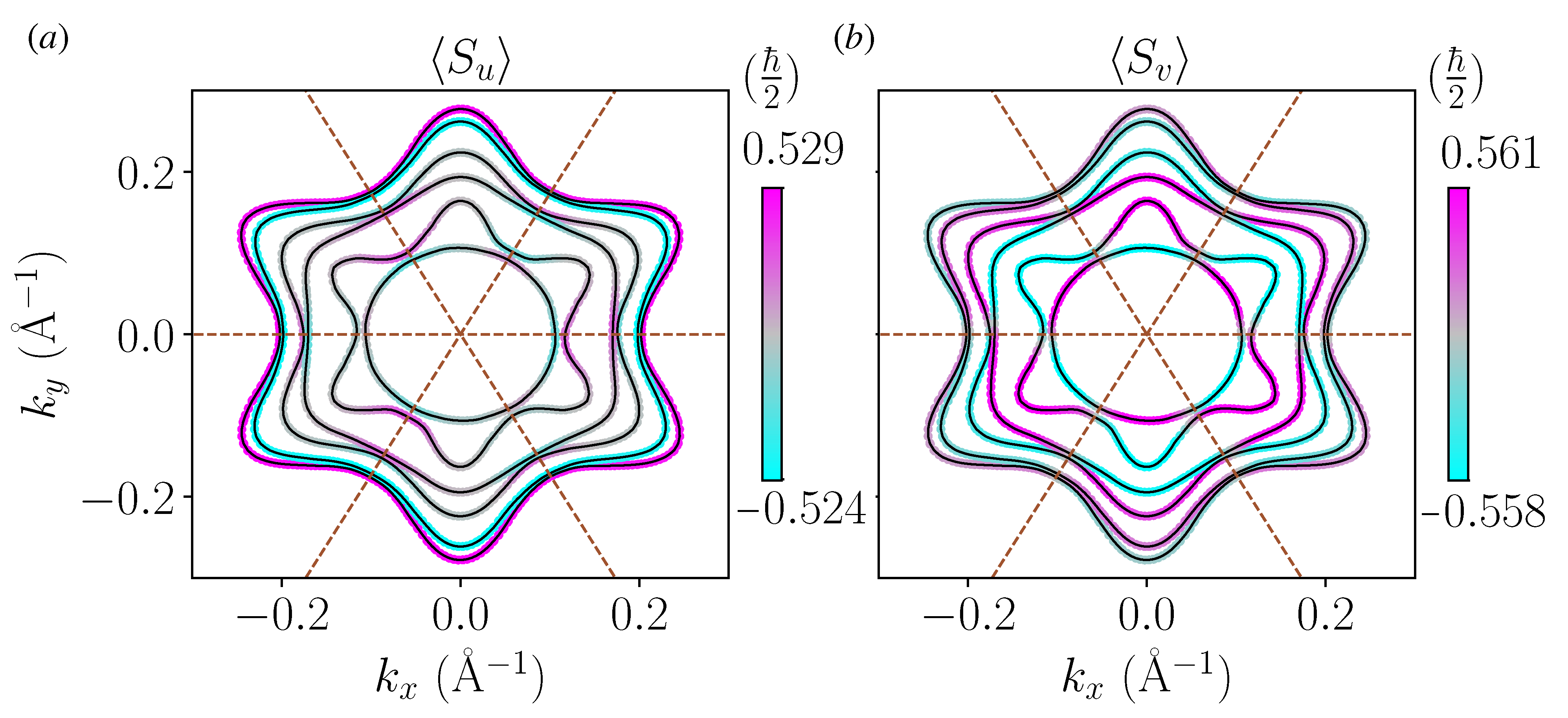}
	\caption{(a-b)Momentum-space $S_u$($S_v$) spin polarized energy iso-surface including spin-orbit coupling of the broken-helical phase in the $k_z=0.2\ \mathbf{b}_3$ plane, for an energy $E=E_F+0.60$ eV. Here, $\mathbf{b}_3$ corresponds to the out-of-plane reciprocal lattice vector.}
 \label{g-wave}
\end{figure}

Finally, for the broken-helical phase, the breaking of the g-wave term together with $[C_{2v}||\mathcal{P}]$, now allows regions with non-zero $S_u$ spin polarization, as shown in Fig. \ref{g-wave}(a). Meanwhile, the g-wave order for $S_v$ remains mostly unchanged under SOC effects, as shown in Fig. \ref{g-wave}(b).

To summarize, we demonstrate that including SOC does not significantly suppress the non-relativistic spin-orderings and spin-splittings observed in both helical phases, suggesting they could potentially be observed experimentally using angle-resolved photoemission spectroscopy (ARPES) \cite{krempasky2024altermagnetic,reimers2024direct} or the magneto-optical Kerr effect technique \cite{gray2024time}. This highlights the importance of using spin symmetries to characterize non-collinear spin arrangements with unconventional magnetic order and to recognize dominant physics arising from exchange.

\section{Linear Edelstein response under SOC effects.}
\label{ed_soc}
\begin{figure}[H]
\centering
	\includegraphics[width=0.5\textwidth]{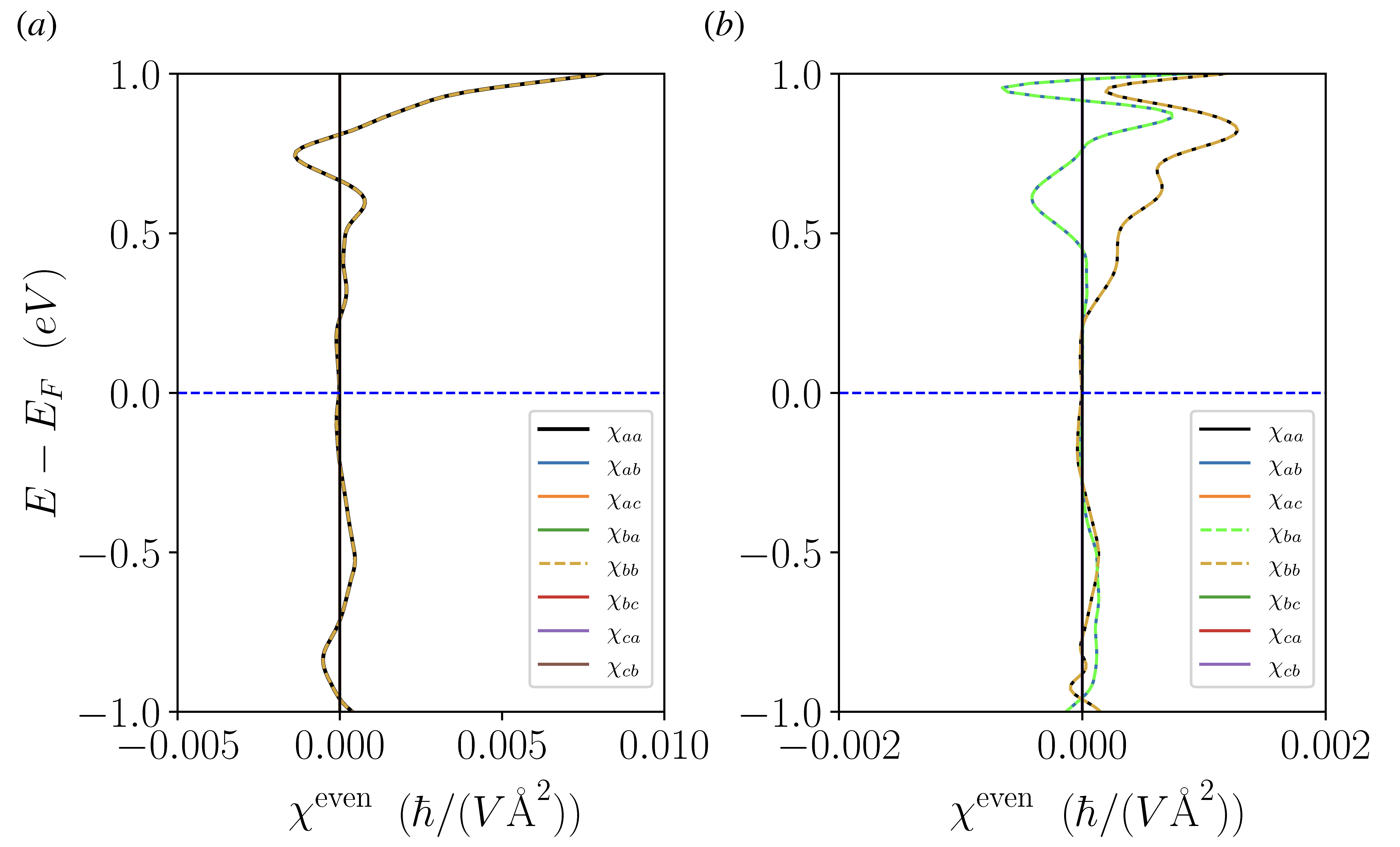}
	\caption{Intraband response tensor $\chi_{i,j}$ including spin-orbit coupling for the (a) helical and (b) broken-helical order, with $\Gamma = 0.01$ eV. All out-of-plane components, except for $\chi_{cc}$ are zero, and the in-plane components adhere to the symmetry constraints outlined in Table \ref{table}.}
 \label{intraband_all}
\end{figure}

As previously discussed, without spin-orbit coupling, the only term that survives the spin symmetry constraints is the out-of-plane Edelstein response $\chi_{cc}$. When spin-orbit coupling is included, an in-plane response emerges for both the $\mathcal{T}$-even and $\mathcal{T}$-odd contributions. Figure \ref{intraband_all} shows the intraband response $\chi_{i,j}$ with SOC for both the helical, and broken-helical order. We see that the in-plane response terms agree with the symmetry constraints listed in Table \ref{table} and are approximately one order of magnitude smaller than the dominant non-relativistic $\chi_{zz}$ term.

\section{Disorder-independent ratio}
\label{cond}

\begin{figure}[H]
\centering
\includegraphics[width=0.49\textwidth]{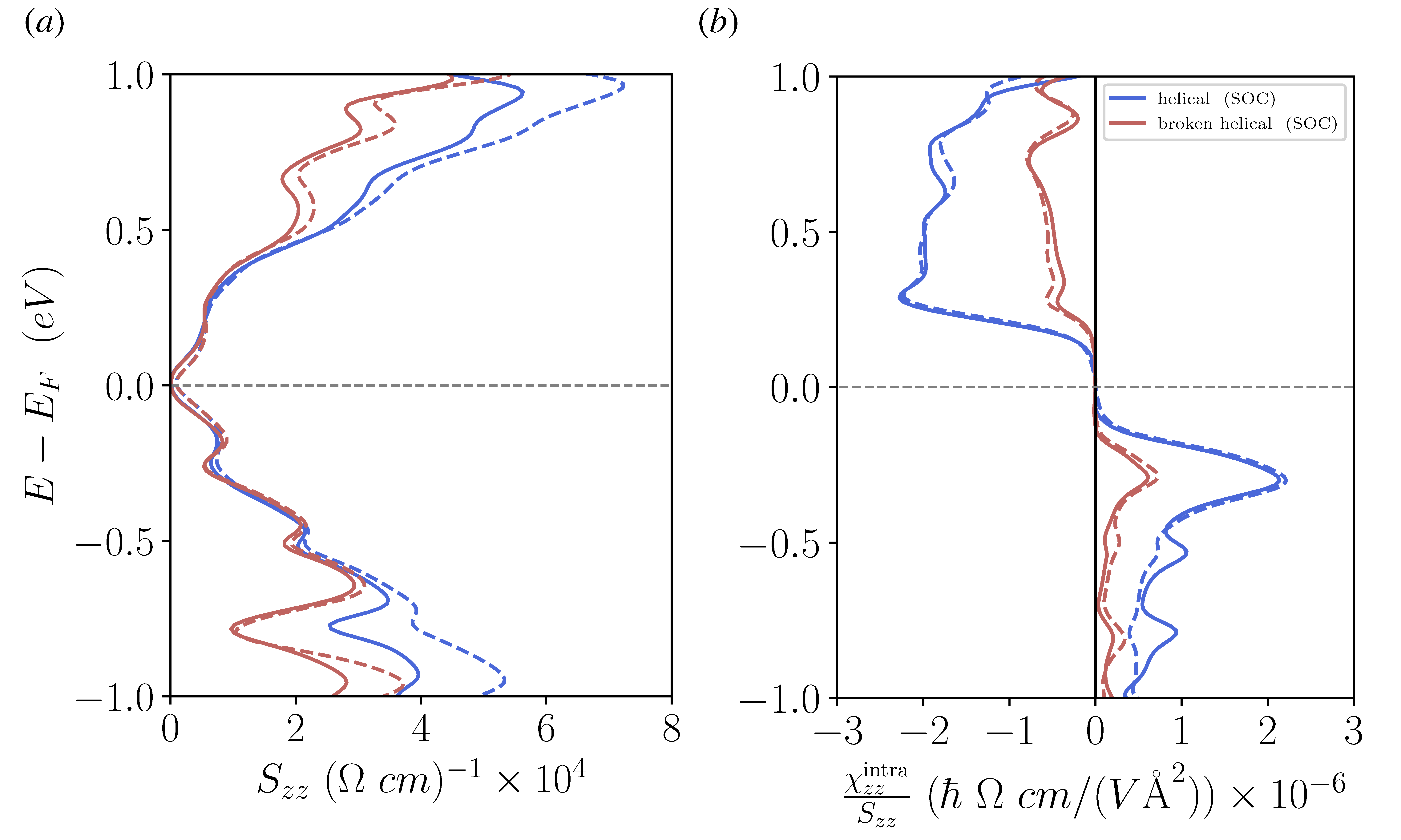}
\caption{(a) Longitudinal $S_{zz}$ conductivity and (b) disorder-independent spin density, taken as the ratio between the intraband spin density and the longitudinal conductivity. The transport calculations, including SOC, for the helical and broken-helical phases correspond to the blue and red solid lines, respectively. The dashed lines represent the calculations without SOC.}
 \label{ratio}
\end{figure}

The intraband spin density scales with disorder scattering as $1/\Gamma$. To isolate the effects of the disorder, we compute the ratio between the intraband response tensor $\chi_{zz}^{intra}$ and the longitudinal conductivity $S_{zz}$ defined in Eq. \ref{lon_cond}. In Fig. \ref{ratio}(a), we show the longitudinal conductivity $S_{zz}$, featuring values within the typical range of conductivities observed in semiconductors. Finally, in Fig \ref{ratio}(b) we plot the disorder-independent ratio, which highlights a clear distinction between both phases, with values in the helical phase four times larger than those in the broken-helical phase.

\begin{equation}
    S_{ij} = \frac{e^2}{\Gamma\hbar}\int \frac{d\mathbf{k}}{(2\pi)^3} \sum_{n} \mathrm{Re}[( \hat{v}_i)_{nn}(\hat{v}_{j})_{nn}] \times \delta({E_{\mathbf{k}n}-E_F})\,,
    \label{lon_cond}
\end{equation}

\bibliography{references}

\end{document}